\documentclass[conference]{IEEEtran}

\usepackage[dvips]{color}
\usepackage{comment}
\usepackage{todonotes}
\usepackage{epsf}
\usepackage{epsfig}
\usepackage{times}
\usepackage{epsfig}
\usepackage{graphicx}
\usepackage{bbold}
\usepackage{mathrsfs}
\usepackage{amssymb}
\usepackage{pdfpages}
\usepackage{epstopdf}
\usepackage{dsfont}
\usepackage{lettrine} 
\usepackage{amsmath,epsfig,amssymb,algorithm,algpseudocode,amsthm,cite,url}
\IEEEoverridecommandlockouts
\usepackage{subcaption}
\allowdisplaybreaks
\usepackage{csquotes}
\usepackage{verbatim}
\usepackage[english]{babel}
\usepackage{amsmath,amssymb}

\usepackage[justification=centering]{caption}
\usepackage{verbatim}

\begin{document}
\clearpage
\title{\huge Resource Allocation for Machine-to-Machine
	Communications with Unmanned Aerial Vehicles}
%
\author{{Mehdi Naderi Soorki$^{1,2}$}, Mohammad Mozaffari$^{1}$, Walid Saad$^{1}$,\\
 Mohammad Hossein~Manshaei$^{2}$, and Hossein~Saidi$^{2}$\vspace*{0.15cm}\\
 \authorblockA{\small $^{1}$Wireless@VT, Bradley Department of Electrical and Computer Engineering, Virginia Tech, Blacksburg, VA, USA,\\ Email: \protect\url{{mehdin,mmozaff,walids}@vt.edu} \\
$^{2}$Department of Electrical and Computer Engineering, Isfahan University of Technology, Isfahan, Iran,\\ Email: \protect\url{{m.naderisoorki}@ec.iut.ac.ir,{manshaei,hsaidi}@cc.iut.ac.ir}\\
}
\vspace{-0.45cm}
  }
%
%
%
%
\maketitle
\thispagestyle{empty}

\begin{abstract}
In this paper, a novel framework for power-efficient, cluster-based machine-to-machine (M2M) communications is proposed. In the studied model, a number of unmanned aerial vehicles (UAVs) are used as aerial base stations to collect data from the cluster heads (CHs) of a set of M2M clusters. To minimize the CHs' transmit power while satisfying the rate requirements of M2M devices, an optimal scheduling and resource allocation mechanism for CH-UAV communications is proposed. First, using the queue rate stability concept, the minimum number of UAVs as well as the dwelling time that each UAV must spend for servicing the CHs are computed. Next, the optimal resource allocation for the CH-UAV communication links is determined such that M2M devices rate requirements are satisfied with a minimum transmit power. Simulation results show that, as the packet transmission probability of machines increases, the minimum number of UAVs required to guarantee the queue rate stability of CHs will also significantly increase. 
Our results also show that, compared  to a case with pre-deployed terrestrial base stations, the average transmit power of CHs will decrease by 68\% when UAVs are used. 
\end{abstract}
\vspace{0.05cm}

\section{Introduction}\vspace{0.1cm}
\label{sec:Intro}
Machine-to-machine (M2M) communications allow the interconnection of a massive number of machine type devices (MTDs) \cite{dawy}. In particular, M2M communications can be used in many Internet of Things (IoT) applications such as intelligent transportation, health care monitoring, smart grid, and public safety \cite{dawy}. To effectively support communications between the massive number of MTDs, a reliable wireless infrastructure is needed. In such M2M scenarios, machine type devices must transmit their data to some existing base stations in the wireless network. However, in areas which experience an intermittent or poor coverage by terrestrial wireless networks, battery-limited MTDs are not able to transmit their data to far away base stations due to their power constraints. Furthermore, due to the various applications of MTDs, they might be deployed in environments with no terrestrial wireless infrastructures such as mountains and desert areas.

In such challenging scenarios, unmanned aerial vehicles (UAVs) can be used as flying base stations to provide reliable and energy-efficient uplink M2M communications \cite{mozaffari2015unmanned,pang,MozaffariIoT}. In particular, UAVs can play a key role in enabling M2M communications when the access to terrestrial wireless networks is limited or unavailable. Due to the aerial nature of the UAVs and their high altitude, they can be effectively deployed to reduce the shadowing and blockage effects  \cite{MozaffariIoT,ZhangLetter,Letter}. Therefore, the UAVs can intelligently move for collecting MTD data. However, to leverage the use of UAVs to collect MTD data, efficient techniques must be developed for resource allocation, network deployment, and multiple access \cite{ghavimi2015m2m, Nof,laya2014}.

Due to the massive number of MTDs, optimizing the resources needed for uplink multiple access becomes highly challenging. In this regard, clustering the M2M devices and employing cluster heads (CHs) for collecting the M2M data and sending it to the aerial base stations, is an effective way to address the massive access problem \cite{tu2011energy,ho2012energy,wei2012joint, Mehrnaz}. In clustered M2M networks, some MTDs can act as cluster heads in order to relay the received packets from the cluster members (CMs) to the base stations. In this case, different criteria such as quality-of-service (QoS), and power consumption of the MTDs can be considered for clustering of the devices \cite{tu2011energy} and~\cite{ho2012energy}. Hence, in a clustered M2M network that is covered by UAVs, the cluster heads are able to send their data to the nearest UAVs with a low transmit power.

The work in \cite{mozaffari2015unmanned} investigated the optimal deployment and trajectory of a single UAV for maximizing downlink coverage. However, the model presented in \cite{mozaffari2015unmanned} does not consider multiple UAVs case. In \cite{pang}, UAVs are used to efficiently collect data from rechargeable CHs. Nevertheless, the work in \cite{pang} does not address the problem of optimal scheduling and resource allocation in UAV-based M2M communications. While some studies such as \cite{ho2012energy} and \cite{azari2015} addressed the problem of M2M scheduling and resource allocation in cellular networks, they do not consider the use of UAVs as aerial base stations in their model. Also, recent works on clustering, such as \cite{MM}, do not incorporate UAVs in their model.  

The main contribution of this paper is to develop a scheduling and resource allocation framework for energy-efficient CH-UAV communications. In particular, we consider a network in which a number of UAVs must provide uplink transmission links to collect the data from the CHs of a number of MTD clusters. Using the queue rate stability concept, we derive the minimum number of required UAVs to serve cluster heads as well as their dwelling time over each CH. Next, considering a minimum rate requirement for CHs, we propose an optimal resource allocation mechanism for CH-UAV communications such that the total transmit power of CHs is minimized. 
Our results show that, the use of UAVs can yield up to a 68\% reduction in the CHs' transmit power as compared to a case with pre-deployed terrestrial base stations. Furthermore, based on the packet transmission probability of the MTDs, our results determine the minimum number of UAVs needed for the rate stability of the CHs' queues.  

The rest of this paper is organized as follows. Section II
presents the system model. In Section III, we present the optimal UAV scheduling model. In Section IV, we introduce the proposed resource allocation mechanism. Simulation results are presented in Section V while conclusions are drawn in Section VI.\vspace{0.01cm}

\section{System Model}\vspace{0.05cm}\label{Sec:Sys-Model}
Consider a number of MTDs distributed over a given geographical area. These MTDs form a set of clusters $\mathcal{G}=\{G_1,G_2,...,G_{|\mathcal{G}|}\}$. In each cluster, an MTD is selected as a CH that is responsible to send the data packets of all of its cluster members using uplink transmission links. Let $\mathcal{I}$ be the set of the indices of CHs. The size of each data packet pertaining to CH $g\in \mathcal{I}$ is $D_g$. In this area, multiple UAVs (at low altitudes) are used as flying base stations to collect data packets from the CHs. We also let $\mathcal{U}$ be the set of $U$ available UAVs. The UAVs must dynamically move and stop over the CHs to collect the uplink data. Clearly, the dwelling time of each UAV over each CH depends on the number of packets that the CH wants to transmit. For multiple access, we consider an orthogonal frequency division multiple access (OFDMA) scheme with $Z$ resource blocks (RBs) each having a bandwidth $B_z$. Let $z_u$ be the number of RBs assigned to a given UAV $u$. For each UAV $u$, we define a vector $\boldsymbol{d}_u=[d_u^g]_{|\mathcal{G}|\times 1}$ with each element being the dwelling time of UAV $u$ needed to cover CH $g$ during $T$. We also define $P_{g,z}$ as the transmit power that a CH $g$ needs for reliable data transmissions over RB $z$.

In Figure 1, we show an illustrative example with three clusters which are being served by two UAVs. In each cluster, there is one CH that must relay the packets of cluster members to its serving UAV during the dwelling time. In this scenario, since the size of cluster $G_2$ is larger than other clusters, $\text{UAV}{_2}$ needs to stay longer over cluster 2. However, the sizes of clusters $G_1$ and $G_3$ are small enough thus allowing $\text{UAV}{_1}$ to change its position and provide service to $G_1$ and $G_3$. Next, we present the queue model for each cluster member. \vspace{0.02cm}
\subsection{Queue of requests}
In each cluster, the queue of data packets at the CH contains all of the data packets received from the CMs. At each time slot, a CM transmits its data packet with a probability $p$ to the CH. Let $a_{g,t}$ be the arrival process of a given data packet to the CH $g$ of cluster $G_g$ during time slot $t$. We assume that the data packets from each CM will immediately trigger an arrival process of data packet to the CH. During each time slot, $a_{g,t}$ can change from $0$ which indicates that none of CMs in $G_g$ transmit data, to $|G_g|$, indicating that all of the MTDs in $G_g$ transmit a data packet to the UAV. The probability with which $a_{g,t}=n$, is given by:
\begin{equation}
\textrm{Pr}(a_{g,t}=n)=\frac{|G_g|!}{n!(|G_g|-n)!}p^n(1-p)^{|G_g|-n}.
\end{equation}
The expected value of $a_{g,t}$, $\bar{a}_{g,t}$, is given by a Binomial distribution as follows:
\begin{equation}
\bar{a}_{g,t}=\sum_{n=0}^{G_g}n\frac{|G_g|!}{n!(|G_g|-n)!}p^n(1-p)^{|G_g|-n}=p|G_g|.
\end{equation}

Let $d_{g,t}$ be the departure process from queue $Q_{g,t}$ at time slot $t$. Note that, multiple UAVs can sequentially cover a CH during each time slot. If the CH $g$ can send at least one packet during the coverage time of UAV $u$, the average departure rate from each queue $g$ is given by:\vspace{-0.2cm}
\begin{equation}
\bar{d}_{g,t}=\frac{\sum_{u=1}^{U} d_g^u(t)}{T}.
\end{equation}

Then, the change in the queue length of cluster $G_g$ will be~\cite{neely2010stochastic}:
\begin{equation}
Q_{g,t+1}=\max\{Q_{g,t}-d_{g,t},0\}+a_{g,t}.\vspace{-0.2cm}
\label{coalition queue}
\end{equation}
\vspace{-0.1cm}
\begin{figure}[!t]
	\begin{center}
		\vspace{-0.2cm}
		\includegraphics[width=3.25in]{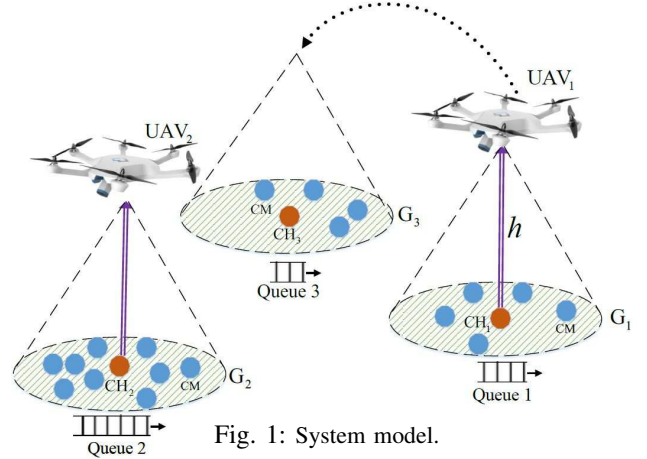}
		\vspace{-0.7cm}
		\caption{ \small System model.}\vspace{-0.5cm}
		\label{Nu}
	\end{center}
\end{figure}

To guarantee that the queue length of CHs does not become infinite, the number of UAVs and their dwelling time for serving the CHs must be sufficient. Therefore, we first find the minimum number of UAVs and their dwelling time over the CHs which ensure the queue rate stability. Next, to establish the successful uplink transmission of CHs, we determine the optimal number of resource blocks and minimum transmit power for each CHs. In this case,  we need to address the problem of optimal resource allocation for energy-efficient CH-UAV communications.
\vspace{0.05cm}
\section{Scheduling and Resource Allocation}\vspace{0.1cm}
Here, we first determine the minimum number of UAVs and their dwelling time to ensure that the queue length of CHs remains bounded over time. Then, we propose an optimal resource allocation (RA) mechanism for the UAVs.
\subsection{UAV Scheduling}
The time duration needed to collect the packets from each CH should be long enough to guarantee the queue
\emph{rate stability}. The rate stability theorem in~\cite{neely2010stochastic} introduces the following rule for stabilizing a multi-queue network: make scheduling decisions such that the average service time and the arrival rates are well defined and satisfy $\bar{a}_{g,t}\leq \bar{d}_{g,t}$ for each queue $g$. Define matrix $\boldsymbol{D}=[\boldsymbol{d}_u]_{U\times|\mathcal{G}|}$, with each column being vector $\boldsymbol{d}_u$. Using this rate stability theorem, the set of scheduling decisions on the service time of UAVs can be written as follows:\vspace{-0.05cm}
\begin{align}\label{opt_problem_UAV.}
\Lambda=\{&\boldsymbol{D}=[\boldsymbol{d}_u]_{U\times|\mathcal{G}|}\mid
\boldsymbol{d}_u,\bar{a}_{g,t}\leq \bar{d}_{g,t}  \text{ , }\forall g\in\mathcal{I},\nonumber\\
&d_u^g\geq 0 \text{, }\forall g\in\mathcal{I},\forall u\in\mathcal{U},\sum_{u} d_u^g=1\text{ , }\forall g\in\mathcal{I}\}.\vspace{-0.3cm}
\end{align}

Based on the definition of $\Lambda$, the minimum number of UAVs is the number for which $\Lambda$ is not empty for the given arrival rates. For example, in Figure 2, two cases are shown for a scenario in which there are two clusters of MTDs. In the first case shown in Figure 2 (a), the average arrival rates of queues in CHs $(a_1,a_2)$ are low enough such that $\Lambda$ is not empty for a serving UAV, which means $\Lambda=\{\boldsymbol{d_1}\mid\boldsymbol{d_1}=
\begin{bmatrix}
d_1^1 \\
d_1^2
\end{bmatrix},d_1^1+d_1^2=1,\bar{a}_{1,t}\leq d_1^1,\bar{a}_{2,t}\leq d_1^2\}$. Thus, one UAV can collect the packets from both CHs and guarantee the rate stability for both queues. In Figure 2 (a), the $x$-axis and $y$-axis, respectively, show the dwelling of UAV 1 on CH 1 and CH 2 for $T=1$. When the average arrival rate of the queues increases from $(a_1,a_2)$ to $(b_1,b_2)$, then, as shown in Figure 2 (a), $\Lambda$ becomes empty for the single UAV case. Therefore, only a single UAV cannot guarantee the rate stability. Based on Figure 2 (b), to obtain a nonempty $\Lambda$, at least two UAVs are needed to collect the packets of each queue. In Figure 2 (b) the $x$-axis and $y$-axis, respectively, show the dwelling time of UAV 1 on CH 1 and UAV 2 on CH 2 for $T=1$.  Then we have $\Lambda=\{\boldsymbol{D}=[\boldsymbol{d_1};\boldsymbol{d_2}]\mid\boldsymbol{d_1}=
\begin{bmatrix}
1,
0
\end{bmatrix}^T,\boldsymbol{d_2}=
\begin{bmatrix}
0,
1
\end{bmatrix}^T\}$.

\begin{figure}[!t]
	\begin{center}
		\vspace{-0.3cm}
		\includegraphics[width=3.35in]{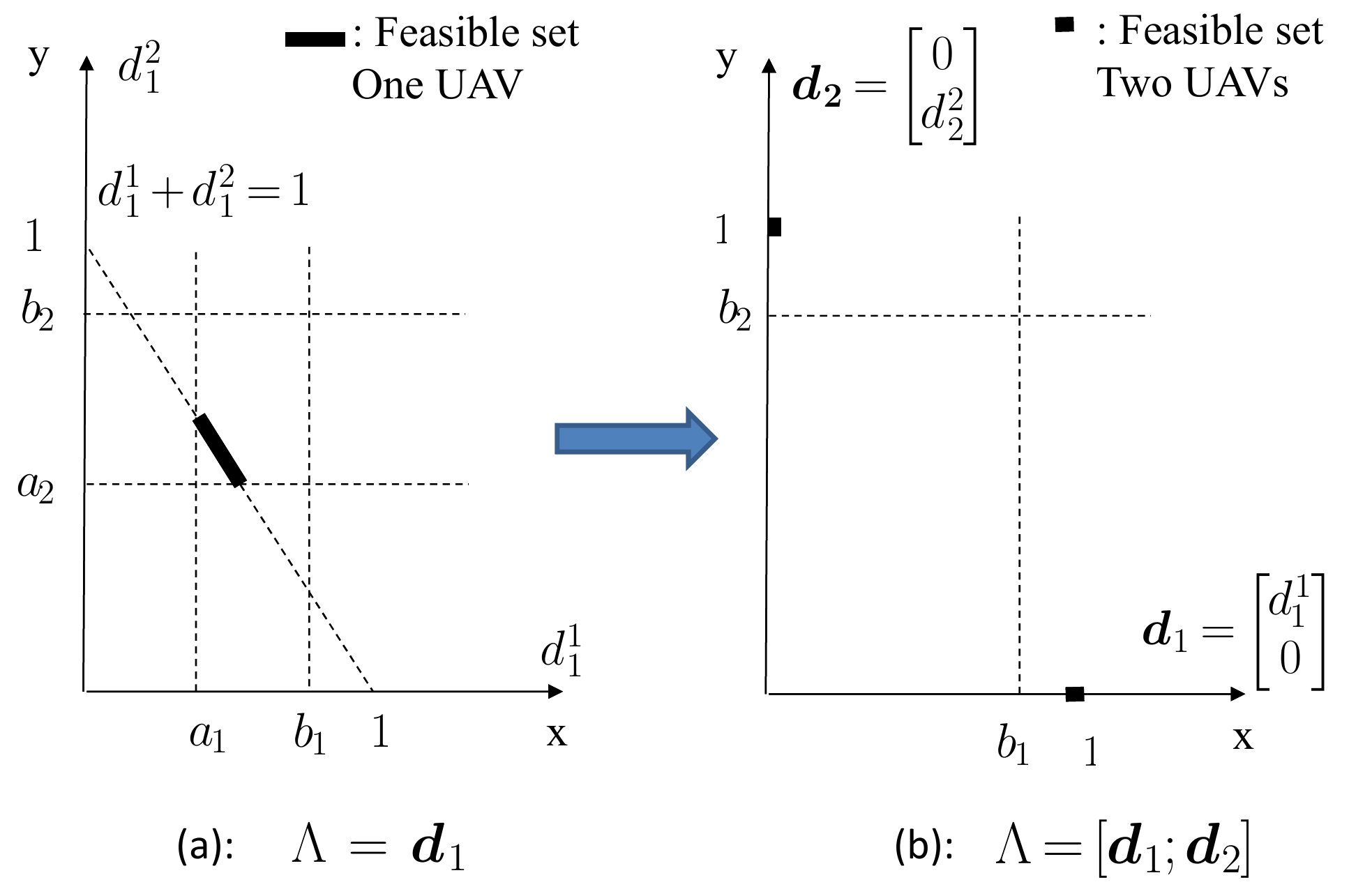}
		\vspace{-0.3cm}
		\caption{ \small Feasible sets and dwelling time of UAVs for two clusters.\vspace{-0.3cm}}\vspace{-0.2cm}
		\label{Nu}
	\end{center}
\end{figure}

Thus far, we have determined the minimum number of UAVs as well as their dwelling time to guarantee the queues' stability of CHs. Next, for each CH, we find the optimal number of resource blocks and minimum transmit power required for the successful and energy-efficient uplink transmissions.

\subsection{Resource Allocation for CHs and UAVs} 
In our model, the system resources include the resource blocks assigned to UAVs and the transmit power of each CHs. Each CH must transmit its packets with a minimum power using the assigned resource blocks. The joint resource block allocation and power control optimization problem that minimizes the total energy consumption of CHs is given by:\vspace{0.2cm}
\begin{align}
&\underset{\boldsymbol{P},\boldsymbol{z}}\min{\sum_g{\sum_u d_g^u \sum_{z_u} P_{g,z}^{u} }},\label{opt_problem_power}\\
&D_g \leq \sum_{z_u} B_z \log\left(1+\frac{P_{g,z}^u \beta H_{gu}^z}{B_zN_0}\right) d_g^u \text{ , }\forall g\in\mathcal{I},\label{opt_problem_power_1}\\
&0<\sum_u z_u \leq Z,\label{opt_problem_power_2}\\
&0<\sum_z{P_{g,z}^u} \leq P_g^{\text{max}}\text{ , }\forall g\in\mathcal{I},\label{opt_problem_power_3}
\end{align}
where $P_{g,z}^u$ is the transmit power of each CH $g$ over sub-channel $z$, when CH $g$ is serviced by UAV $u$. $\boldsymbol{P}=[P_{g,z}^u]_{|\mathcal{G}|\times U\times Z}$ is a matrix with each element being the power $P_{g,z}^u$. $|\mathcal{G}|$ is the number of CHs, $U$ is the number of UAVs, and $Z$ is the number of RBs. The vector $\boldsymbol{z}=[z_u]_{U\times 1}$ indicates the number of resource blocks assigned to UAV $u$. $P_g^u=\sum_{z_u}P_{g,z}^u$ is the sum of the transmit power of CH $g$ over all of the RBs assigned to UAV $u$. $H_{gu}^z$ is the channel gain between UAV $u$ and CH $g$ on each sub-channel $z$, and $\beta=\frac{-1.5}{\ln(5P_e)}$ is the SNR gap for M-QAM modulation with $P_e$ being the maximum acceptable error probability~\cite{goldsmith2005wireless}. Here, we have $H_{gu}^z={(\frac{4\pi h_{gu}}{\lambda})^{-\nu}}$ where $\nu$ is the path loss exponent, $h_{gu}$ is the distance of UAV $u$ from cluster head $g$, and $\lambda$ is the wavelength. $P_g^{\text{max}}$ and $Z$ are, respectively, the maximum transmit power of CH $g$ and the maximum number of available RBs.

In the defined optimization problem (\ref{opt_problem_power}), constraint (\ref{opt_problem_power_1}) guarantees that each CH can send one packet during the time that the UAV covers it. Also, (\ref{opt_problem_power_2}) indicates the maximum number of resource blocks available for UAVs. (\ref{opt_problem_power_3}) shows the maximum transmission power of CHs. In general, solving (\ref{opt_problem_power}) is challenging due to the presence of both integer, and continuous, variables as well as a nonlinear constraint (\ref{opt_problem_power_1}). Thus, we relax this problem by assuming that each CH will use the same transmit power on all sub-carriers. Subsequently, considering $P_{g,z}^u=\frac{P_g^u}{z_u}$, the optimization problem in (\ref{opt_problem_power}) can be reformulated as:\vspace{0.2cm}
\begin{align}
&\underset{\boldsymbol{P},\boldsymbol{z}}\min{\sum_gP_g^u{\sum_ud_g^u }},\label{altered_opt_problem_power}\\
& B_zN_0(2^{\frac{D_g}{z_uB_z d_g^u}}-1)\frac{{z_u}}{\beta H_{gu}}\leq {P_g^u}  \text{ , }\forall g\in\mathcal{I},\label{altered_opt_problem_power_1}\\
&\sum_u z_u\leq Z,\label{altered_opt_problem_power_2} \\
& 0< z_u\leq Z\text{ , }\forall u \in\mathcal{U},\label{altered_opt_problem_power_3}\\
&0<{P_g^u} \leq P_g^{\text{max}}\text{ , }\forall g \in\mathcal{I},\label{altered_opt_problem_power_4}
\end{align}
where each element of the matrix $\boldsymbol{P}=[P_g^{u}]_{|\mathcal{G}|\times U}$ indicates the transmit power of each CH $g$, when it is serviced by UAV $u$. Clearly, the optimization problem in ($\ref{altered_opt_problem_power}$) is a mixed integer programming problem. We relax this optimization problem by considering continuous values for variable $z_u$. Obviously, the objective function in ($\ref{altered_opt_problem_power}$) is linear, and, hence, it is convex. Furthermore, considering the fact that the function $(2^{\frac{1}{x}}-1)x-y$ is a convex function on $(x,y)\in \mathds{R}^2$, constraint (\ref{altered_opt_problem_power_1}) is also convex as it is a sub-level set of a convex function on $z_u$ and $P_g^u$. Constraints (\ref{altered_opt_problem_power_2}) and (\ref{altered_opt_problem_power_3}) are also the sub-level sets of convex functions. Consequently, the optimization problem in (\ref{altered_opt_problem_power}) is a convex optimization problem.

Since the optimization problem in (\ref{altered_opt_problem_power}) is convex, the strong duality will hold. Therefore, any pair of primal and dual optimal points must satisfy the Karush-Kuhn-Tucker (KKT) conditions~\cite{boyd2004convex}. Following the KKT conditions, the Lagrange multipliers and optimal points are given by:\vspace{0.1cm}
\begin{align}
&B_zN_0(2^{\frac{D_g}{z_u^* B_z d_g^u}}-1)\frac{{z_u^*}}{\beta H_{gu}}\leq {P_g^{u,*}}  \text{ , }\forall g\in\mathcal{I},\forall u \in\mathcal{U},\nonumber\\
&0< z_u^*\leq Z\text{ , }\forall u \in\mathcal{U},\nonumber\\
&0<{P_g^{u,*}} \leq P_g^{\text{max}}\text{ , }\forall g\in\mathcal{I},\nonumber\\
&\sum_u z_u^*\leq Z,\nonumber\\
&\lambda_1^{u,*},\lambda_2^{u,*},\lambda_1^{g,*},\lambda_2^{u,*},\lambda_6^{g,u,*}\geq0\text{ , }\forall g\in\mathcal{I},\forall u\in\mathcal{U},\nonumber\\
&\lambda_5^*\geq0 ,\nonumber\\
&{\sum_ud_g^u }+\lambda_1^{g,*}+\lambda_2^{g,*}-\lambda_6^{g,u,*}=0\text{ , }\forall g\in\mathcal{I},\forall u\in\mathcal{U},\nonumber\\
&-\lambda_1^{u,*}-\lambda_2^{u,*}+\lambda_5^{*}+\nonumber\\
&\lambda_6^{g,u,*}\times\frac{B_zN_0} {\beta H_{gu}}\left(2^{\frac{D_g}{z_u^*B_zd_g^u}}-\frac{2^{\frac{D_g}{z_u^*B_zd_g^u}}\log(2)}{z_u^*}-1\right)=0 \, , \nonumber\\
&\lambda_1^{u,*}\times(-z_u^*)=0\text{ , }\forall u \in\mathcal{U},\nonumber\\
&\lambda_2^{u,*}\times(z_u^*-Z)=0\text{ , }\forall u \in\mathcal{U},\nonumber\\
&\lambda_1^{g,*}\times(-p_g^*)=0\text{ , }\forall g\in\mathcal{I},\nonumber\\
&\lambda_2^{g,*}\times(p_g^*-p_g^{\text{max}})=0\text{ , }\forall g\in\mathcal{I},\nonumber\\
&\lambda_5^{*}\times(\sum_u z_u^*-Z)=0 \, ,\nonumber\\
&\lambda_6^{g,u,*}\times \left(B_zN_0(2^{\frac{D_g}{z_u^*B_z d_g^u}}-1)\frac{{z_u^*}}{\beta H_{gu}}-{P_g^{u,*}}\right)=0\text{ , }\forall g\in\mathcal{I},\nonumber
\end{align}
where $[{\lambda_1^{u}}^*]_{U\times 1},{\lambda_2^{u}}^*]_{U\times 1},[{\lambda_1^{g}}^*]_{|\mathcal{G}|\times 1},[{\lambda_2^{g}}^*]_{|\mathcal{G}|\times 1},{\lambda_5}^*$ and $[{\lambda_6^{g,u}}^*]_{|\mathcal{G}|\times U}$ are the Lagrange multipliers. Since $ z_u^*>0$ and ${P_g^{u,*}}>0$ following $\lambda_1^{g,*}\times(p_g^*)=0$ and $\lambda_1^{u,*}\times(-z_u^*)=0$, both  $\lambda_1^{g,*}$ and $\lambda_1^{u,*}$ must be zero. Thus, we consider $\lambda_1^{g,*}=0$ and $\lambda_1^{u,*}=0$ in the KKT conditions. The feasibility conditions for the optimization variables in problem (\ref{altered_opt_problem_power}) will be:
\begin{align}
&B_zN_0(2^{\frac{D_g}{z_u^* B_z d_g^u}}-1)\frac{{z_u^*}}{\beta H_{gu}}\leq {P_g^{u,*}}  \text{ , }\forall g\in\mathcal{I},\label{f_lagrange_s}\\
&0< z_u^*\leq Z\text{ , }\forall u\in\mathcal{U},\\
&0<{P_g^{u,*}} \leq P_g^{\text{max}}\text{ , }\forall g\in\mathcal{I},\\
&\sum_u z_u^*\leq Z,\\
&\lambda_2^{u,*},\lambda_2^{u,*},\lambda_6^{g,u,*}\geq0\text{ , }\forall g\in\mathcal{I},\forall u\in\mathcal{U},\\
&\lambda_5^*\geq0.\label{f_lagrange_e}
\end{align}

Therefore, considering (\ref{f_lagrange_s})-(\ref{f_lagrange_e}), the optimal Lagrange multipliers, $[{\lambda_2^{u}}^*]_{U\times 1},[{\lambda_2^{g}}^*]_{|\mathcal{G}|\times 1},{\lambda_5}^*$ and $[{\lambda_6^{g,u}}^*]_{|\mathcal{G}|\times U}$, optimal power of cluster head $g$ to transmit packet to UAV $u$, $[{P_g^{u}}^*]_{|\mathcal{G}|\times U}$, and optimal number resource block for each UAV $u$ $[{z_u^*}]_{U\times 1}$ are given by the following joint nonlinear equations:
\begin{align}
&{\lambda_2^{u}}^*\times(Z-z_u^*)=0\text{ , }\forall u \in\mathcal{U},\label{optimal points_s}\\
&{\lambda_2^{g}}^*\times(p_g^*-p_g^{\text{max}})=0\text{ , }\forall g\in\mathcal{I},\\
&{\lambda_5}^{*}\times(\sum_u z_u^*-Z)=0\text{ , }\forall u\in\mathcal{U},\\
&{\sum_ud_g^u}+{\lambda_2^{g}}^*-{\lambda_6^{g,u}}^*=0,\\
&-{\lambda_2^{u}}^*+{\lambda_5}^*+\\
&{\lambda_6^{g,u}}^*\times\frac{B_zN_0} {\beta H_{gu}}\left(2^{\frac{D_g}{z_u^*B_zd_g^u}}-\frac{2^{\frac{D_g}{z_u^*B_zd_g^u}}\log(2)}{z_u^*}-1\right)=0\text{ , }\\
&{\lambda_6^{g,u}}^*\times \left(B_zN_0(2^{\frac{D_g}{z_u^*B_z d_g^u}}-1)\frac{{z_u^*}}{\beta H_{gu}}-{{P_g^{u}}^*}\right)=0\text{ , }\nonumber\\
&\forall g\in \mathcal{I}, \, \forall u\in\mathcal{U}.
\label{optimal points_n}
\end{align}

To solve the joint nonlinear equations (\ref{optimal points_s})-(\ref{optimal points_n}), we use the Levenberg Marquardt algorithm (LMA)~\cite{gavin}. This algorithm requires an initial point (as an input) for the variables in the feasible set of constraints (\ref{f_lagrange_s})-(\ref{f_lagrange_e}). Typically, LMA is able to find the optimal solution even with initial points that are far from the optimal one. As a result of solving (\ref{optimal points_s})-(\ref{optimal points_n}), the optimal number of resource blocks as well as the transmit power of each cluster head are computed. \vspace{0.15cm}

\section{Simulation Results}\vspace{0.2cm}

For our simulations, we consider a set of $|\mathcal{G}|$ clusters uniformly distributed in  a square geographical area 500\,m$\times$500\,m. The number of cluster members randomly changes between 1 to 10. The bandwidth of each resource block is $15$~KHz. We consider a $2$~GHz carrier frequency and a maximum power of $1$\,W for each CH. This power is equally divided among all resource blocks. The noise power spectral density $N_0$ is $-170$~dBm per Hz. Furthermore, due to flight regulations and environmental obstacles around the locations of CHs, we assume that the altitudes of the UAVs randomly change from $400$\,m to $600$\,m. We consider a path loss exponent of $2.5$ for the CHs-UAVs communications. Moreover, the length of each packet is set to $100$~bits, and the target bit error rate is $10^{-7}$. All statistical results are averaged over a large number of independent runs.

\begin{figure}[t]
\centering
\includegraphics[width=3.15in]{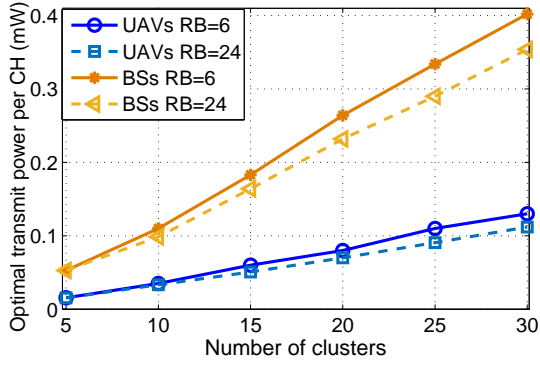}\vspace{-0.2cm}
\caption{\small Average optimal transmit power per CHs vs. number of clusters for 0.1 packet transmission probability.}
\vspace{-0.5cm}
\label{Optimum_power}
\end{figure}
In Figure \ref{Optimum_power}, we show the average transmit power of CHs versus the number of clusters for two scenarios. In the first scenario, we use multiple UAVs as aerial base stations, and in the second scenario, multiple terrestrial base stations (equal to the number of UAVs), are uniformly deployed for serving the CHs. As we can see from Figure \ref{Optimum_power}, using aerial base stations leads to about 68\% power reduction for CHs compared to terrestrial base stations. For instance, for 20 clusters and 6 RBs, the transmit power of CHs decreases from 0.26\,mW to 0.09\,mW by using UAVs instead of terrestrial base stations. In fact, the UAVs can effectively move towards the CHs and significantly reduce the blockage and shadowing effects. Hence, the CHs can use lower transmit power for sending their data to the UAVs than the ground base stations. From Figure \ref{Optimum_power}, we can see that the average transmit power per CH increases as the number of clusters increases. This is due to the fact that, when the number of CHs increases, the average time that each UAV can spend to collect data from each CH will decrease. Thus, the CH must send its packet with a minimum time duration. Consequently, the CH increases its transmission power to meet the target packet rate during the given short time duration. Clearly, while satisfying the rate requirements, the transmit power of CHs can be reduced by increasing the number of resource blocks. Therefore, in Figure \ref{Optimum_power}, the average transmit power of CHs for 24 RBs is lower than the one for 6 RBs.

\begin{figure}[t]
	\centering
	\includegraphics[width=3.1in]{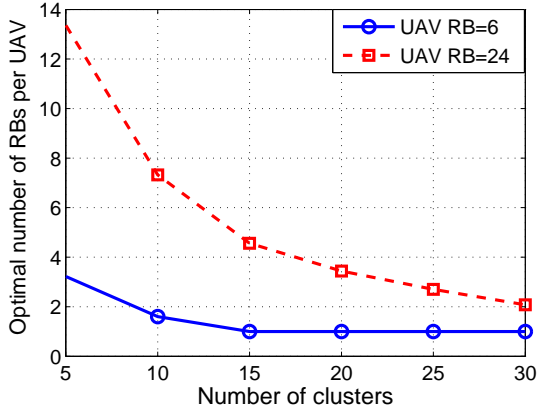}\vspace{-0.3cm}
	\caption{\small Average number of RB per UAVs vs. number of clusters when the probability of packet transmission is $0.1$.}
	\label{optimal_RB}
	\vspace{-0.4cm}
\end{figure}

In Figure \ref{optimal_RB}, we show the impact of the number of clusters on the average number of resource blocks that must be assigned to each UAV. From Figure \ref{optimal_RB}, we can see that the average number of resource blocks per UAV is lower for a higher number of clusters. This is due to the fact that, as the number of clusters increases, the minimum number of needed UAVs will increase. Hence, the number of RBs per UAV decreases accordingly. For instance, as shown in Figure \ref{optimal_RB}, as the number of clusters increases from 10 to 20, the average number of RBs per UAV decreases from 7 to 3, for a total of 24 RBs. 

\begin{figure}[t]
\centering
\includegraphics[width=3.1in]{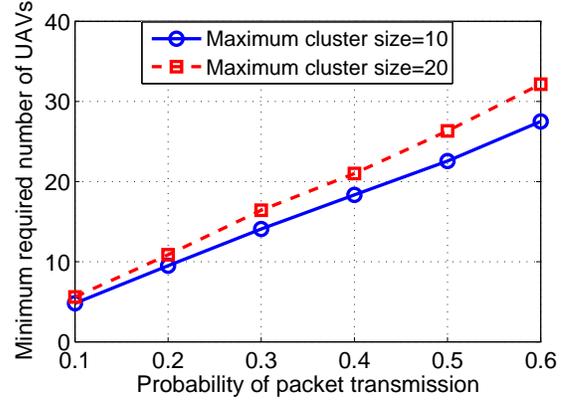}
\caption{\small Minimum required number of UAVs vs. the probability of packet transmission for a scenario with 12 RBs.}
\vspace{-0.3cm}
\label{Min_Number_UAV}
\end{figure}
Figure \ref{Min_Number_UAV} shows the impact of packet transmission probability on the minimum required number of UAVs. From Figure \ref{Min_Number_UAV}, we can see that the minimum required number of UAVs significantly increases when the probability of packet transmission increases. This is due to the fact that, for a higher packet transmission probability, the average arrival rate of packets to the cluster heads will increase. Therefore, the minimum required number of UAV to guarantee the rate stability must be increased. Moreover, as seen from Figure \ref{Min_Number_UAV}, the minimum required number of UAVs is also affected by the cluster size. Clearly, more UAVs are needed to serve larger clusters. For instance, for a packet transmission probability of 0.5, 4 additional UAVs are needed when the cluster size increases from 10 to 20.

\begin{figure}[!t]
	\begin{center}
		\vspace{-0.2cm}
		\includegraphics[width=8.3cm]{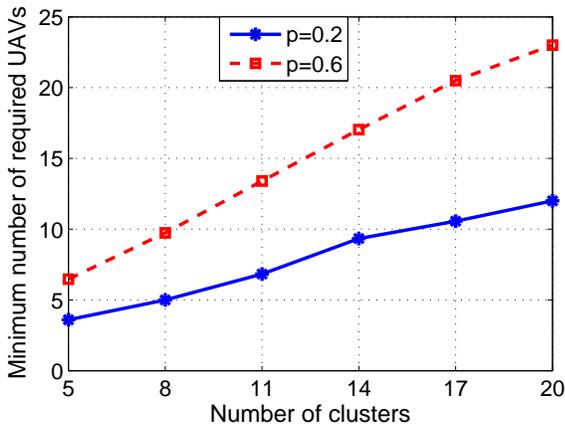}
		\vspace{-0.01cm}
		\caption{ \small Minimum number of required UAVs vs. number of clusters. \vspace{-0.3cm}}\vspace{-0.1cm}
		\label{UAV_number}
	\end{center}
\end{figure}

In Figure \ref{UAV_number}, we show how the number of UAVs required for serving the CHs changes as a function of the number of clusters. From this figure, we can see that, as the number of clusters increases, more UAVs must be deployed in order to successfully serve the cluster heads.   
 For example, according to Figure \ref{UAV_number}, when the number of clusters grows from 5 to 20,  the number of UAVs must increase from 6 to 23 for $p=0.6$. Furthermore, comparing the slope of the curves for $p=0.2$ and $p=0.6$, indicates that  the rate of increase in the number of UAVs versus the number of clusters is higher for a larger transmission probability.

\begin{figure}[!t]
\begin{center}
\vspace{-0.2cm}
\includegraphics[width=9.0cm]{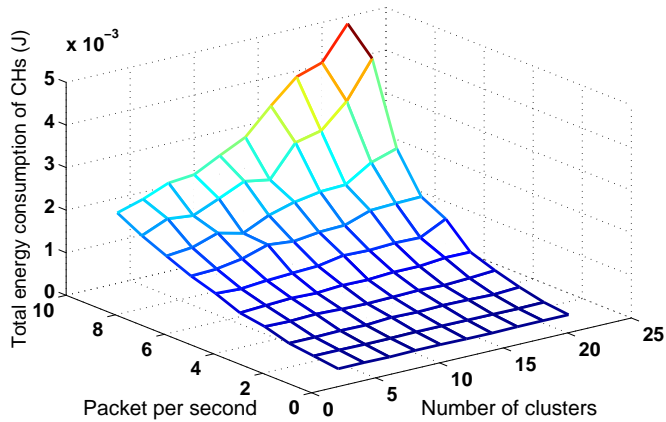}
\vspace{-0.5cm}
\caption{ \small The total energy consumption of CHs vs. target packet rate and number of CHs, when number of RB is $12$.\vspace{-0.3cm}}
\label{sum_energy}\vspace{-0.2cm}
\end{center}
\end{figure}
Figure \ref{sum_energy} shows the total energy consumption of CHs as a function of the target packet rate and the number of CHs. From Figure \ref{sum_energy}, we can see that the total energy consumption of CHs increases as the target packet rate or the number of CHs increases. For higher target packet rate values, the cluster heads must transmit more bits within the given time duration. Therefore, the CHs must increase their transmit power to send more packets during the presence of the UAVs. Moreover, as the number of clusters increases, the UAVs must decrease their dwelling time over each cluster head in order to be able to serve all the cluster heads. Consequently, given the shorter dwelling time, each CH will increase its transmit power to send the data to the serving UAV.   \vspace{0.2cm} 

\section{Conclusions}\label{Sec:Conclusion}
In this paper, we have proposed a novel framework for power-efficient cluster-based M2M communications. In the studied model, given clusters of M2M devices, UAVs are used as aerial base stations to collect data from the cluster heads (CHs). In this scenario, we have proposed an optimal scheduling and resource allocation mechanism for CHs-UAVs communications to minimize the transmit power of CHs while satisfying M2M devices rate requirements. To this end, first, using the queue rate stability concept, we have computed the minimum number of UAVs, and the dwelling time of each UAV to cover CHs. Next, we have determined the optimal resource allocation for CHs-UAVs communications such that M2M devices rate requirements are met with a minimum transmit power. Simulation results have shown the various benefits and tradeoffs of using UAVs to service M2M clusters.\vspace{-0.05cm}

\bibliographystyle{IEEEtran}
\def\baselinestretch{1.04}
\bibliography{references}

\begin{thebibliography}{10}
\providecommand{\url}[1]{#1}
\csname url@samestyle\endcsname
\providecommand{\newblock}{\relax}
\providecommand{\bibinfo}[2]{#2}
\providecommand{\BIBentrySTDinterwordspacing}{\spaceskip=0pt\relax}
\providecommand{\BIBentryALTinterwordstretchfactor}{4}
\providecommand{\BIBentryALTinterwordspacing}{\spaceskip=\fontdimen2\font plus
\BIBentryALTinterwordstretchfactor\fontdimen3\font minus
  \fontdimen4\font\relax}
\providecommand{\BIBforeignlanguage}[2]{{%
\expandafter\ifx\csname l@#1\endcsname\relax
\typeout{** WARNING: IEEEtran.bst: No hyphenation pattern has been}%
\typeout{** loaded for the language `#1'. Using the pattern for}%
\typeout{** the default language instead.}%
\else
\language=\csname l@#1\endcsname
\fi
#2}}
\providecommand{\BIBdecl}{\relax}
\BIBdecl

\bibitem{dawy}
Z.~Dawy, W.~Saad, A.~Ghosh, J.~G. Andrews, and E.~Yaacoub, ``Towards massive
  machine type cellular communications,'' \emph{IEEE Wireless Communications
  Magazine, to appear}, 2016.

\bibitem{mozaffari2015unmanned}
M.~Mozaffari, W.~Saad, M.~Bennis, and M.~Debbah, ``{Unmanned aerial vehicle
  with underlaid device-to-device communications: performance and tradeoffs},''
  \emph{IEEE Transactions on Wireless Communications}, vol.~15, no.~6, pp.
  3949--3963, June. 2016.

\bibitem{pang}
Y.~Pang, Y.~Zhang, Y.~Gu, M.~Pan, Z.~Han, and P.~Li, ``Efficient data
  collection for wireless rechargeable sensor clusters in harsh terrains using
  {UAV}s,'' in \emph{Proc. of IEEE Global Communications Conference
  (GLOBECOM)}, Austin, TX, USA, Dec. 2014, pp. 234--239.

\bibitem{MozaffariIoT}
M.~Mozaffari, W.~Saad, M.~Bennis, and M.~Debbah, ``Mobile {Internet of Things}:
  Can {UAVs} provide an energy-efficient mobile architecture?'' in \emph{Proc.
  of IEEE Global Communications Conference (GLOBECOM)}, Washington, DC, USA,
  Dec. 2016.

\bibitem{ZhangLetter}
J.~Lyu, Y.~Zeng, and R.~Zhang, ``Cyclical multiple access in {UAV}-aided
  communications: A throughput-delay tradeoff,'' \emph{available online:
  arxiv.org/abs/1608.03180}, 2016.

\bibitem{Letter}
M.~Mozaffari, W.~Saad, M.~Bennis, and M.~Debbah, ``Efficient deployment of
  multiple unmanned aerial vehicles for optimal wireless coverage,'' \emph{IEEE
  Communications Letters}, vol.~20, no.~8, pp. 1647--1650, Aug. 2016.

\bibitem{ghavimi2015m2m}
F.~Ghavimi and H.-H. Chen, ``{M2M} communications in {3GPP} {LTE/LTE-A}
  networks: Architectures, service requirements, challenges, and
  applications,'' \emph{IEEE Communications Surveys \& Tutorials}, vol.~17,
  no.~2, pp. 525--549, Oct. 2015.

\bibitem{Nof}
N.~Abuzainab, W.~Saad, and H.~V. Poor, ``Cognitive hierarchy theory for
  heterogeneous uplink multiple access in the {Internet of Things},'' in
  \emph{Proc. of IEEE International Symposium on Information Theory {(ISIT)}},
  Barcelona, Spain, June. 2016.

\bibitem{laya2014}
A.~Laya, L.~Alonso, and J.~Alonso-Zarate, ``Is the random access channel of
  {LTE} and {LTE-A} suitable for {M2M} communications? a survey of
  alternatives,'' \emph{IEEE Communications Surveys \& Tutorials}, vol.~16,
  no.~1, pp. 4--16, Dec. 2014.

\bibitem{tu2011energy}
C.-Y. Tu, C.-Y. Ho, and C.-Y. Huang, ``Energy-efficient algorithms and
  evaluations for massive access management in cellular based machine to
  machine communications,'' in \emph{Proc. of IEEE Vehicular Technology
  Conference (VTC Fall)}, San Francisco, CA, Sept. 2011, pp. 1--5.

\bibitem{ho2012energy}
C.~Y. Ho and C.-Y. Huang, ``Energy-saving massive access control and resource
  allocation schemes for {M2M} communications in {OFDMA} cellular networks,''
  \emph{IEEE Wireless Communications Letters}, vol.~1, no.~3, pp. 209--212,
  2012.

\bibitem{wei2012joint}
S.-E. Wei, H.-Y. Hsieh, and H.-J. Su, ``Joint optimization of cluster formation
  and power control for interference-limited machine-to-machine
  communications,'' in \emph{Proc. of IEEE Global Communications Conference
  (GLOBECOM)}, Anaheim, CA, Dec. 2012, pp. 5512--5518.

\bibitem{Mehrnaz}
M.~Afshang, H.~S. Dhillon, and P.~H.~J. Chong, ``Coverage and area spectral
  efficiency of clustered device-to-device networks,'' in \emph{Proc. of IEEE
  Global Communications Conference (GLOBECOM)}, San Diego, USA, Dec. 2015.

\bibitem{azari2015}
A.~Azari and G.~Miao, ``Lifetime-aware scheduling and power control for {M2M}
  communications in {LTE} networks,'' in \emph{Proc. of IEEE Vehicular
  Technology Conference (VTC)}, Glasgow, May. 2015, pp. 1--5.

\bibitem{MM}
T.~Wang, L.~Song, Z.~Han, and W.~Saad, ``Distributed cooperative sensing in
  cognitive radio networks: An overlapping coalition formation approach,''
  \emph{IEEE Transactions on Communications}, vol.~62, no.~9, pp. 3144--3160,
  Sept. 2014.

\bibitem{neely2010stochastic}
M.~J. Neely, ``{Stochastic network optimization with application to
  communication and queueing systems},'' \emph{Synthesis Lectures on
  Communication Networks}, vol.~3, no.~1, pp. 1--211, 2010.

\bibitem{goldsmith2005wireless}
A.~Goldsmith, \emph{Wireless communications}.\hskip 1em plus 0.5em minus
  0.4em\relax Cambridge university press, 2005.

\bibitem{boyd2004convex}
S.~Boyd and L.~Vandenberghe, \emph{Convex optimization}.\hskip 1em plus 0.5em
  minus 0.4em\relax Cambridge university press, 2004.

\bibitem{gavin}
H.~Gavin, ``The levenberg-marquardt method for nonlinear least squares
  curve-fitting problems,'' \emph{Department of Civil and Environmental
  Engineering, Duke University}, pp. 1--15, 2011.

\end{thebibliography}
\end{document}